\documentclass[prb,aps,twocolumn,amsmath,amssymb,superscriptaddress,longbibliography]{revtex4-2}
\usepackage{graphicx} 
\usepackage{float}
\usepackage{amsmath}
\usepackage{amssymb}
\usepackage{amsfonts}
\usepackage{euscript}
\usepackage{enumerate}
\usepackage{hhline}
\usepackage{pslatex}
\usepackage{tabularx}
\usepackage{braket}
\usepackage{soul}
\usepackage{bibunits}
\usepackage[normalem]{ulem}
\usepackage[dvipsnames]{xcolor}

\newcommand{\ysnote}[1]{{\color{red} #1}}

\makeatletter
\renewcommand{\@biblabel}[1]{#1. }
\renewcommand{\@dotsep}{500}
\renewcommand{\@pnumwidth}{0em}
\renewcommand{\l@figure}[2]{
\newcommand{\ysnote}[1]{{\color{red} #1}}
\@dottedtocline{1}{1.5em}{2em}{Figure #1}{}\vspace{15pt}}
\usepackage[colorlinks,linkcolor=red,anchorcolor=blue,citecolor=blue]{hyperref}
\begin{document}

\title{Topological Edge States Emerging from Twisted Moiré Bands}

\author{Yasser Saleem}
\affiliation{
Institute for Theoretical Physics and Astrophysics,
and Würzburg-Dresden Cluster of Excellence on Complexity, Topology and Dynamics in Quantum Matter ctd.qmat,
Julius-Maximilians-Universität Würzburg, Am Hubland, D-97074 Würzburg, Germany}
\author{Pawe\l ~Potasz}
\affiliation{Institute of Physics, Faculty of Physics, Astronomy and Informatics, Nicolaus Copernicus University, Grudziadzka 5, 87-100 Toru\'n, Poland}
\author{Anna Dyrda\l}
\affiliation{
Faculty of Physics, Adam Mickiewicz University in Pozna\'n}

\author{Björn Trauzettel}
\affiliation{
Institute for Theoretical Physics and Astrophysics,
and Würzburg-Dresden Cluster of Excellence on Complexity, Topology and Dynamics in Quantum Matter ctd.qmat,
Julius-Maximilians-Universität Würzburg, Am Hubland, D-97074 Würzburg, Germany}
\author{Ewelina M. Hankiewicz}
\affiliation{
Institute for Theoretical Physics and Astrophysics,
and Würzburg-Dresden Cluster of Excellence on Complexity, Topology and Dynamics in Quantum Matter ctd.qmat,
Julius-Maximilians-Universität Würzburg, Am Hubland, D-97074 Würzburg, Germany}

\date{\today}
\begin{abstract}
We study twisted bilayer WSe$_2$ within a continuum moir\'e model and apply a method for treating finite geometries directly in the continuum framework, thereby avoiding the limitations associated with purely momentum-space formulations and Wannier obstructions. By projecting a confinement potential onto bulk moir\'e eigenstates, we obtain a real-space description of edge physics without lattice models. Applying this approach to nanoribbons, we demonstrate chiral edge modes consistent with bulk Chern numbers and reveal their moir\'e-scale character. In the magic-angle regime, these states are strongly localized, exhibit layer-polarized counter-propagating modes, and are electrically tunable via a displacement field, enabling control of localization, hybridization, and topological transitions. Our results establish a general framework for boundary physics in topological moir\'e materials.
\end{abstract}

\maketitle
\section{Introduction}
The emergence of topological electronic bands in moir\'e superlattices has opened a powerful route toward engineering quantum phases of matter in van der Waals heterostructures~\cite{Bistritzer2011moire,Cao2018,Cao2018SC,Wu20219ContinuumModel,Devakul2021Twisted,li2026quantum}. In these systems, a small twist angle produces a long-wavelength moir\'e potential that can strongly suppress the electronic kinetic energy, thereby enhancing the role of Coulomb interactions and giving rise to narrow, isolated minibands. As a consequence, moiré materials provide a natural setting for the interplay of topology and strong correlations.

Among the various moiré structures, twisted bilayer graphene initially established this field, but semiconducting transition-metal dichalcogenide (TMD) bilayers have rapidly
emerged as a particularly attractive alternative due to their
strong spin–orbit coupling, valley-contrasting band structure, and the large effective masses of their quasiparticles~\cite{Xiao2012,Xu2014}. In twisted TMD homobilayers, these effects are especially pronounced, as the moiré superlattice generates extremely narrow and well-isolated bands near the valence-band edge, with bandwidths of only a few meV or less~\cite{naik2018ultraflatbands,Wu2018,Devakul2021Twisted}. These features make TMD moiré systems a particularly clean and tunable platform for realizing interaction-driven many-body phases. Experiments have revealed a broad family of emergent states arising from these flat bands, including correlated Mott-like insulators, charge-ordered and generalized Wigner crystal phases at commensurate fillings~\cite{Regan2020,Tang2020,wang2020correlated,zhang2021electronic}, as well as superconductivity in systems such as twisted bilayer WSe$_2$\cite{xia2025superconductivity,guo2025superconductivity}. In related twisted TMD platforms, particularly MoTe$_2$, even more exotic regimes have been observed, including integer and fractional quantum anomalous Hall states, fractional Chern insulators, fractional quantum spin Hall phases, and signatures of composite Fermi liquids\cite{cai2023signatures,park2023observation,zeng2023thermodynamic,xu2023observation,kang2024evidence,ji2024local,redekop2024direct,anderson2024trion,wang2025hidden,xu2025interplay,xu2025signatures,park2025ferromagnetism,thompson2025microscopic,deng2025nonmonotonic,li2025universal,pan2026optical,huber2026optical,park2026observation,holtzmann2026optical,li2026signatures,chen2026visualizing,kang2025time,wang2026magnetic,sun2026twist,kang2024double}.

Both WSe$_2$ and MoTe$_2$ are of particular interest. Their low-energy valence states reside near the $\pm K$ valleys and are subject to strong spin-valley locking, so that each valley sector can be treated as effectively spin polarized at low energies~\cite{Xiao2012,Xu2014}. These twisted homobilayers are commonly described by valley-resolved continuum models in which the moir\'e potential determines the electronic structure near the valence-band edge~\cite{Wu20219ContinuumModel,Devakul2021Twisted,christos2025approximate}. For suitable parameters in the small-angle regime, a single spin-valley sector can be mapped onto a two-band Haldane model, for which the two uppermost bands carry opposite Chern numbers, $(+1,-1)$~\cite{Wu20219ContinuumModel,Devakul2021Twisted,haldane1988model}. At larger twist angles, particularly in WSe$_2$, the upper bands can instead carry same-sign Chern numbers, $(+1,+1)$, requiring a three-band description beyond the two-band Haldane mapping. In the large-angle limit, this corresponds to the triangular $\pi/3$-Hofstadter regime~\cite{crepel2024bridging}. Twisted TMDs thus provide a promising route to realizing topological flat bands and effective lattice models related to the Haldane and Kane--Mele paradigms~\cite{haldane1988model,kane2005quantum,kane2005z}.

From the theoretical side, continuum descriptions of twisted TMD homobilayers established that moir\'e minibands can acquire nontrivial topology due to the combined effects of interlayer hybridization, spin-valley locking, and symmetry-allowed moir\'e potentials~\cite{Devakul2021Twisted,Wu20219ContinuumModel,pan2020band,jia2024moire}. In particular, near special twist angles, the top valence moir\'e band of twisted WSe$_2$ can become nearly flat while retaining substantial and relatively uniform Berry curvature~\cite{Devakul2021Twisted,jia2024moire}.

An important feature of these moir\'e bands is that their topology is encoded in the Berry-curvature distribution across the moir\'e Brillouin zone and in the associated topological invariants, such as valley Chern numbers~\cite{thouless1983quantization,hatsugai1993chern}. Since the two valleys carry opposite Berry curvature by time-reversal symmetry, interaction-driven valley polarization can convert valley-contrasting topology into experimentally observable anomalous Hall responses. At the same time, the topological character of isolated moir\'e bands may obstruct the construction of symmetric, exponentially localized Wannier functions~\cite{thouless1984wannier,brouder2007exponential,po2018fragile}, complicating the derivation of minimal lattice descriptions and raising fundamental questions about the proper low-energy degrees of freedom.

The existence of isolated Chern bands also implies, through the bulk--boundary correspondence, the presence of chiral edge modes in finite geometries~\cite{thouless1983quantization,hatsugai1993chern}. However, while the bulk topology and correlated phases of twisted TMD homobilayers have been studied extensively within continuum and effective tight-binding descriptions, the structure of their edge states remains comparatively unexplored~\cite{shayeganfar2024terahertz,an2017realization,he2026multiple}. Finite-size geometries have been analyzed theoretically in twisted bilayer graphene~\cite{fleischmann2018moire,fujimoto2021moire,wang2023twisted,sanchez2024edge,wania2024atomistic}, but such approaches often rely on atomistic methods that are substantially more tractable in graphene than in twisted TMDs because of the multi-orbital character of the low-energy states. In contrast, continuum moir\'e models for twisted TMDs are naturally formulated in momentum space and rely on translational symmetry, rendering the implementation of physical boundaries nontrivial. Moreover, the Wannier obstruction associated with isolated Chern bands precludes strictly localized single-band real-space descriptions~\cite{thouless1984wannier,brouder2007exponential,po2018fragile}, thereby necessitating multiband formulations. These challenges have likely hindered a systematic theoretical understanding of edge-state structure in twisted TMD homobilayers. Nevertheless, recent experiments have begun to probe one-dimensional boundary and domain-wall modes in moir\'e TMD systems, particularly in twisted bilayer WTe$_2$, where tuning the interlayer coupling via twist provides a means to engineer topology~\cite{lupke2022quantum}.

Recent experiments have established a pronounced twist-angle evolution of correlated phases in twisted bilayer WSe$_2$. Across the intermediate- and larger-angle range, $2.1^\circ\lesssim\theta\leq5.0^\circ$, Xia et al. and Guo et al. mapped correlated insulating and superconducting regimes and showed how the superconducting phase evolves with twist angle and bandwidth~\cite{xia2026bandwidth,guo2026angle}. In the small-angle regime, zero-field Chern-insulating states were observed near $\theta=1.23^\circ$, while direct evidence for a spontaneous $C=1$ quantum anomalous Hall phase was subsequently reported at $\theta=2^\circ$~\cite{foutty2024mapping,gao2026probing}. Theoretical studies associate the small-angle regime with valley-polarized Chern ferromagnetism and propose antiferromagnetic and superconducting states at larger angles~\cite{crepel2024bridging,munozsegovia2025twist,christos2025approximate,fischer2025theory}.

 Against this experimental and theoretical background, several issues remain open for twisted bilayer WSe$_2$ in the presence of boundaries. While the bulk band topology, including the emergence of isolated Chern bands, has been extensively analyzed within continuum models, comparatively less is known about how these features manifest at edges. In particular, the structure and robustness of edge states, their dependence on twist angle, displacement field, and boundary geometry, and their connection to bulk topological invariants and effective tight-binding descriptions require further clarification. A systematic analysis of boundary effects is therefore an important step toward understanding the broader many-body phase diagram of the material.

In this work, we investigate the topological properties of twisted bilayer WSe$_2$ ribbons within a continuum moir\'e model incorporating spin-valley locking, layer-dependent moir\'e potentials, and interlayer tunneling. While such models have been highly successful in describing bulk band topology, their formulation in momentum space makes the treatment of physical boundaries nontrivial. Here we address this limitation by applying a bulk-state projection approach to finite geometries within the continuum framework. By projecting a confinement potential onto a truncated basis of bulk moir\'e eigenstates, we construct an effective real-space description that captures edge physics without relying on Wannierization.

Applying this approach to twisted WSe$_2$ nanoribbons, we demonstrate the emergence of chiral edge modes consistent with the bulk Chern numbers, and reveal that these states possess a distinctly moir\'e-scale character. In contrast to conventional models such as the Bernevig–Hughes–Zhang (BHZ) model\cite{bernevig2006quantum}, where nontrivial topology arises from band inversion between electron- and heavy-hole–like subbands, the topology in twisted WSe$_2$ originates from the moiré-modulated electronic structure, with edge states inheriting a strong layer (pseudospin) and valley character. In particular, near the magic-angle regime where the top valence band is maximally flat, the edge modes are strongly localized on the moir\'e length scale, exhibit pronounced layer (pseudospin) polarization, and display minimal penetration into the bulk. We further show that a perpendicular displacement field provides continuous and highly tunable control over their spatial profile, layer polarization, and hybridization with bulk states. Our results establish a direct connection between continuum moir\'e band topology and boundary-state engineering, and provide a general framework for studying edge phenomena in topological moir\'e materials beyond the limitations of lattice-based descriptions.

This paper is organized as follows. In Sec.\ref{sec:model}, we introduce the continuum model for twisted WSe$_2$ and present our approach to finite nanoribbon geometries, and analyze their electronic properties as a function of twist, with particular focus on the magic angle. In Sec.\ref{sec:efield}, we investigate the effect of a perpendicular displacement field and demonstrate electrical control of the edge-state properties and topology. In Sec.\ref{sec:Haldane}, we compare our continuum results to an effective Haldane model description. Finally, in Sec.\ref{sec:Disc}, we summarize our findings and outline possible directions for future work.

\section{Model For Moiré Nanoribbons}
\label{sec:model}
\begin{figure}[t]
    \centering
    \includegraphics[width=\linewidth]{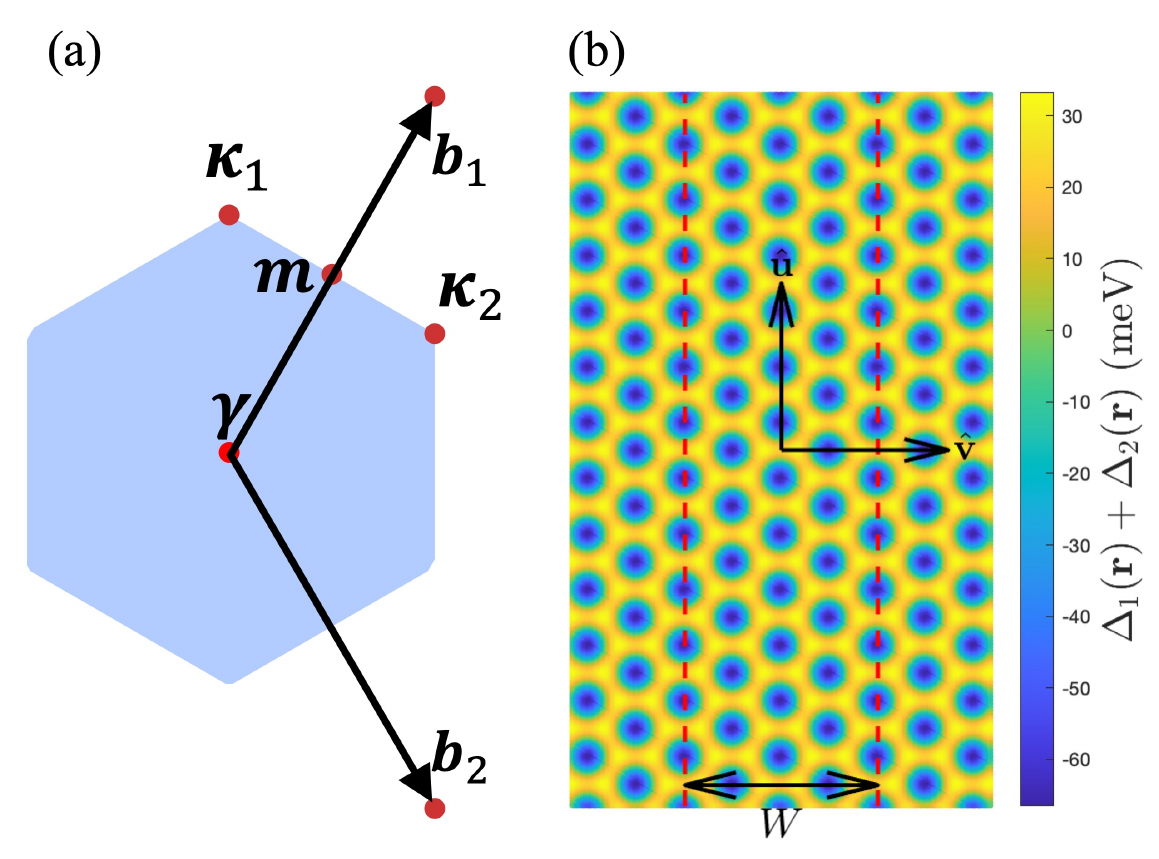}
    \caption{
     Moiré Brillouin-zone geometry and real-space moiré potential for twisted
$\mathrm{WSe_2}$ at a twist angle $\theta = 1.43^\circ$.
(a) Schematic of the moiré Brillouin zone showing the high-symmetry points
$\boldsymbol{\gamma}$, $\boldsymbol{m}$, and $\boldsymbol{\kappa}_{1,2}$, together with the reciprocal lattice vectors $\mathbf{b}_1$ and $\mathbf{b}_2$.
(b)  Real-space moiré potential $\Delta_1(\mathbf{r}) + \Delta_2(\mathbf{r})$ defined by Eq.~(\ref{eq: MoirePotential}) plotted over several moiré periods. The nanoribbon geometry is defined within the region between the red dashed lines, where an additional confinement potential $V(\mathbf{r})$ is applied, while the surrounding area represents the bulk system. The orientation is rotated for visual clarity and the schematic is not to scale.
    }
    \label{fig:fig1}
\end{figure}
\begin{figure*}[t]
    \centering
    \includegraphics[width=\linewidth]{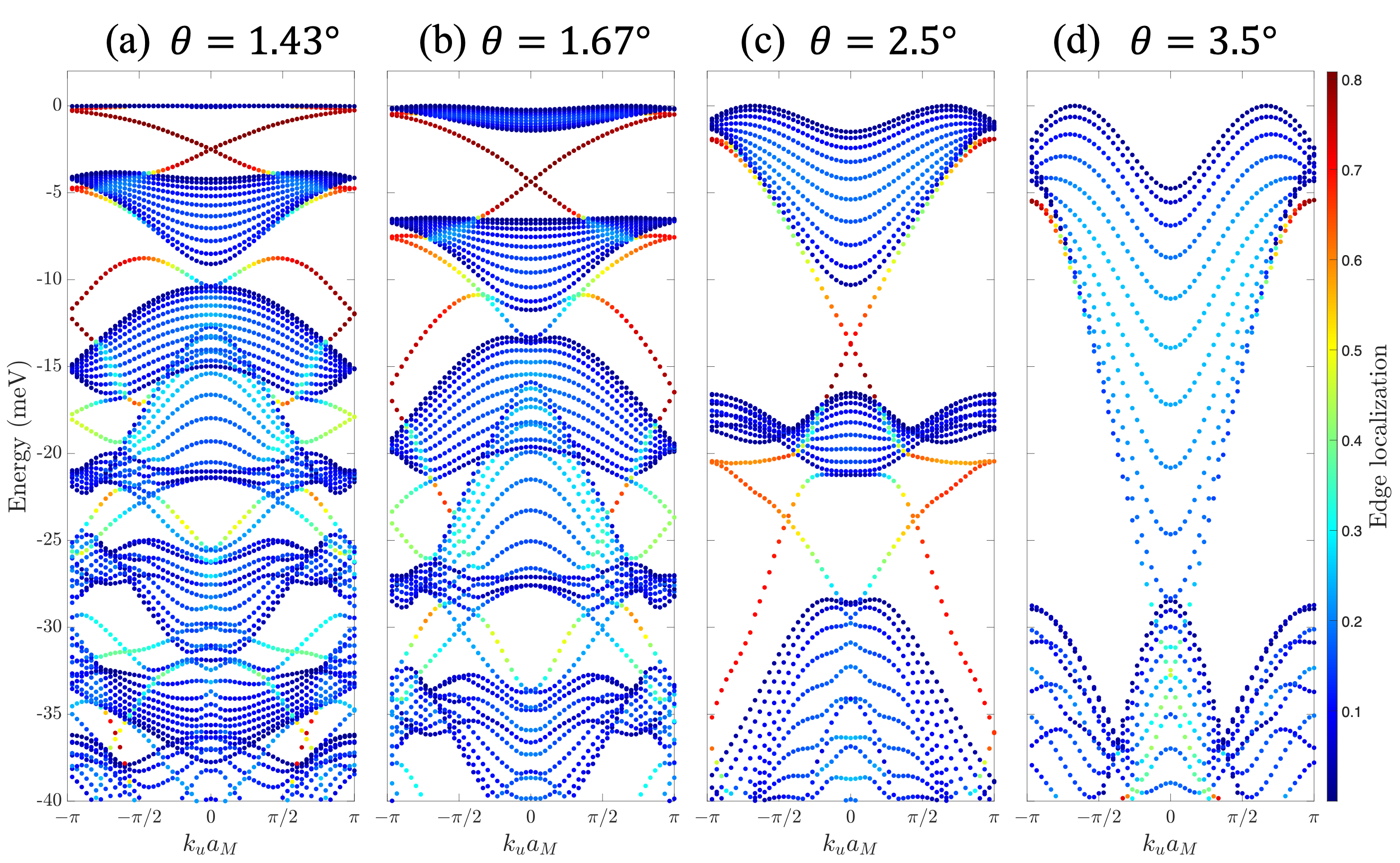}
    \caption{
        Band structures for t$\mathrm{WSe_2}$ zigzag edged nanoribbons for different twist angles $\theta$. Energies are plotted as a function of the moiré quasi-momentum $k_u a_M$ and shifted such that the highest occupied band lies at zero energy. The color scale encodes the degree of edge localization of each eigenstate, with warmer colors indicating stronger localization at the ribbon edges.  
    }
    \label{fig:fig2}
\end{figure*}
\begin{figure}[t]
    \centering
    \includegraphics[width=\linewidth]{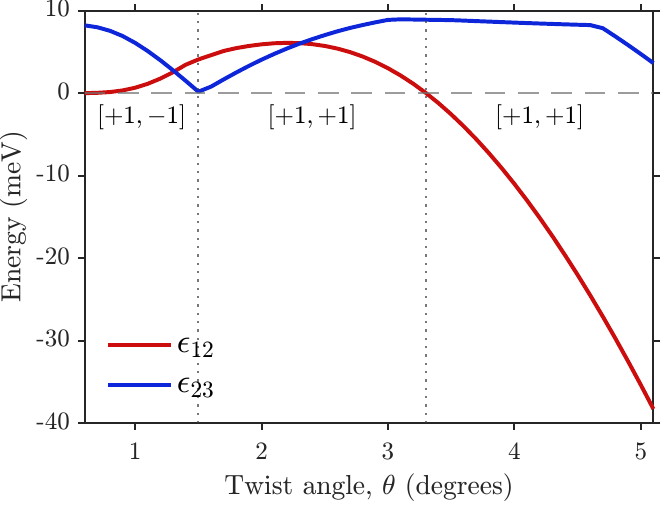}
    \caption{
    Energy gaps of the bulk continuum model bands for twisted WSe$_2$ as a function of twist angle. 
We plot $\epsilon_{12}=\min(E_1)-\max(E_2)$ and $\epsilon_{23}=\min(E_2)-\max(E_3)$, where $E_1$, $E_2$, and $E_3$ denote the first, second, and third valence moir\'e bands, respectively. 
The vertical dotted line at angles $\theta\approx1.5^\circ$ and $\theta\approx3.3^\circ$ mark the gap closings separating the Chern-number regimes indicated in the figure.
    }
    \label{fig:fig3}
\end{figure}
\begin{figure*}[t]
    \centering
    \includegraphics[width=\linewidth]{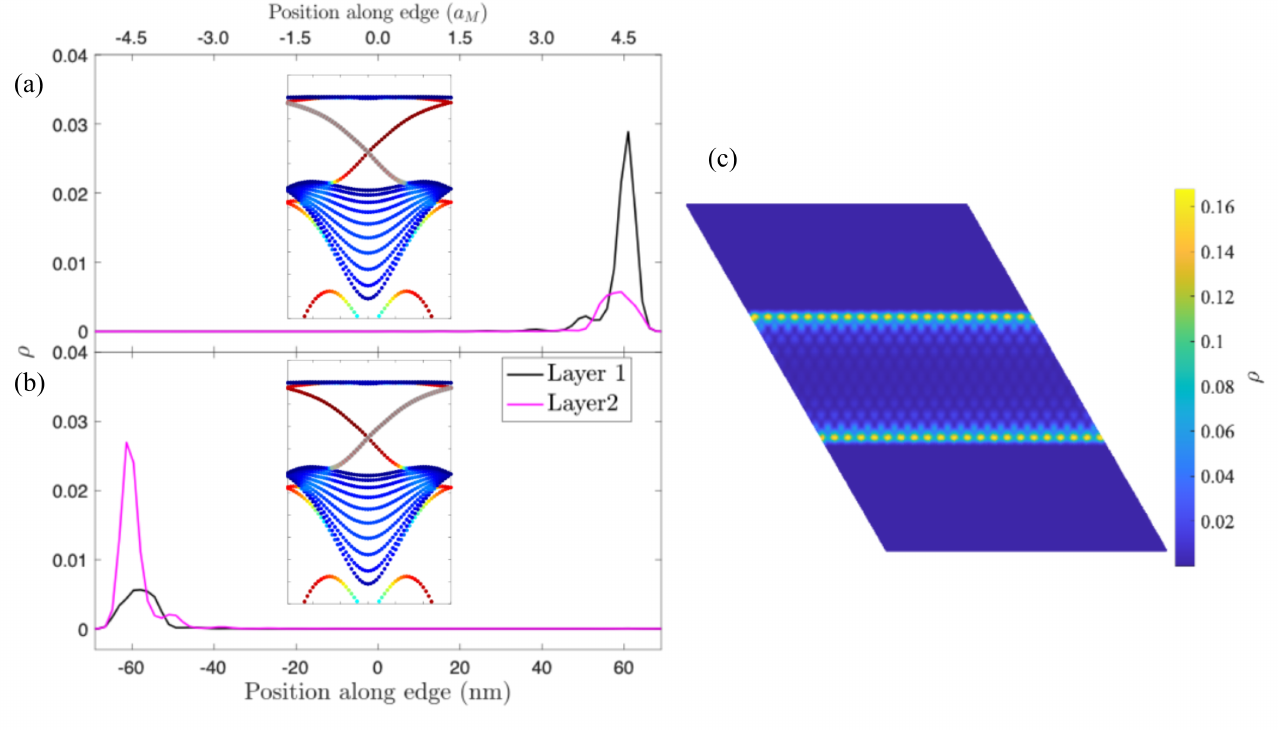}
    \caption{
     (a,b) Real-space charge density profiles of representative edge states for a twisted $\mathrm{WSe_2}$ (t$\mathrm{WSe_2}$) nanoribbon at twist angle $\theta = 1.43^\circ$. The charge density $\rho$, integrated across the ribbon width, is plotted as a function of position along the edge. black and magenta curves denote contributions from layer 1 and layer 2, respectively. Panel (a) corresponds to the right-moving edge mode, while panel (b) shows the left-moving (counter-propagating) mode. Insets display the corresponding edge-state dispersions, with the selected branch highlighted in gray. (c) Spatial map of the total real-space charge density $\rho$ obtained by summing the contributions of the right- and left-moving edge states shown in (a) and (b).
    }
    \label{fig:fig4}
\end{figure*}
\noindent We first consider modeling the bulk. Due to the large unit cells generated by small twist angles,
first-principle density functional theory (DFT) simulations of twisted TMD homobilayers are computationally demanding. Consequently, we adopt the continuum model introduced by Wu \textit{et al.}~[\onlinecite{Wu20219ContinuumModel}]. to describe AA-stacked twisted $\mathrm{WSe_2}$. The continuum model is based on an effective mass approximation of electrons moving through a periodic moiré potential. We take the moiré lattice vectors as $\textbf{a}_1=a_M\left(\frac{\sqrt{3}}{2},-\frac{1}{2}\right) $ and $\textbf{a}_2=a_M\left(\frac{\sqrt{3}}{2},\frac{1}{2}\right)$, where the moiré lattice length for a twist angle $\theta$ is defined by $a_M=a_0/\theta$ with $a_0=0.3317$ nm being the lattice constant for $\mathrm{WSe_2}$. The moiré reciprocal lattice vectors are defined by $\textbf{a}_i\cdot\textbf{b}_j=2\pi \delta _ {ij}$. We take $N_1(N_2)$ unit cells along the $\textbf{a}_1(\textbf{a}_2)$ directions and apply periodic boundary conditions. The Hamiltonian is given by 
\begin{equation}
H_\uparrow =
\begin{pmatrix}
-\dfrac{\hbar^2}{2m^\ast}\lvert \mathbf{k}-\boldsymbol{\kappa}_1 \rvert^2 + \Delta_1(\mathbf{r}) & \Delta_T(\mathbf{r}) \\[8pt]
\Delta_T^\dagger(\mathbf{r}) & -\dfrac{\hbar^2}{2m^\ast}\lvert \mathbf{k}-\boldsymbol{\kappa}_2 \rvert^2 + \Delta_2(\mathbf{r})
\label{eq:ContinHam}
\end{pmatrix}
\end{equation}
$H_\downarrow$ is related by time reversal symmetry (TRS).
$\kappa_{1,2}$ corresponds to two high symmetry points of the moiré Brillouin zone (BZ) and are shown in Fig.~\ref{fig:fig1}(a). The layer dependent moiré potential $\Delta_{1,2}(\mathbf{r})$ is given by
\begin{equation}
\Delta_{1,2}(\mathbf{r})=2V\sum_{j=1,3,5}\cos{\left(\textbf{g}_j\cdot\textbf{r}\pm\varphi\right)},
\label{eq: MoirePotential}
\end{equation}
where we take the plus sign for layer 1, and the minus sign for layer 2 and $\textbf{g}_j$ are $\frac{(j-1)\pi}{3}$ rotations of the reciprocal lattice vector $\textbf{b}_2$.  The inter-layer tunelling
$\Delta_T(\mathbf{r})$ is
\begin{equation}
\Delta_T(\mathbf{r}) = w \left( 1 + e^{-i\mathbf{g}_2 \cdot \mathbf{r}} + e^{-i\mathbf{g}_3 \cdot \mathbf{r}} \right).
\end{equation}
The $\mathrm{WSe_2}$  homobilayer parameters from Ref.~[\onlinecite{Devakul2021Twisted}] are $V=9$ meV, $\varphi=128^\circ$,$w=18$ meV and $m*=0.43m_e$ where $m_e$ is the bare electron mass. We note that this first-harmonic parametrization is not unique, particularly near the magic-angle regime where lattice relaxation can modify miniband gaps and Chern numbers. Relaxation-aware continuum models can yield different Chern-number sequences for the upper moir\'e bands~\cite{zhang2025twist}; therefore, the detailed twist-angle phase sequence discussed below should be understood as specific to the parametrization used here. The eigenstates of the continuum Hamiltonian in Eq.~(\ref{eq:ContinHam}) are expanded in a plane-wave basis of moiré reciprocal lattice vectors,
\begin{equation}
    \Phi_{p,\mathbf{k}}(\mathbf{r})
    =
    \frac{1}{\sqrt{A}}
    \sum_{\mathbf{G}}
    e^{i(\mathbf{k}+\mathbf{G})\cdot\mathbf{r}}
    \begin{pmatrix}
        A^{1}_{p,\mathbf{k}+\mathbf{G}} \\
        A^{2}_{p,\mathbf{k}+\mathbf{G}}
    \end{pmatrix},
    \label{eq: BulkWF}
\end{equation}
where $p$ labels the band index, $\mathbf{k}$ is the crystal momentum within the moiré Brillouin zone, and \(\mathbf{G}\) are moiré reciprocal lattice vectors. The corresponding bulk eigenenergies and eigenstates are obtained from
$
H_{\uparrow}\Phi_{p,\mathbf{k}}(\mathbf{r}) = \epsilon_{p,\mathbf{k}}\Phi_{p,\mathbf{k}}(\mathbf{r}).
$
Here $A$  denotes the total area of the finite moiré supercell.

Our goal is to study confined topological edge states derived from the bulk moiré bands, focusing in particular on hole confinement. To do so, we consider a hard-wall confinement geometry that effectively realizes a moiré nanoribbon, as illustrated in Fig.~\ref{fig:fig1}(b), although the formalism can be straightforwardly generalized to other confinement profiles. Similar approaches have been employed previously to study confined states in bilayer graphene systems\cite{Saleem2023Nano,Korkusinski2023Nanoletters,Sadecka2024PRB,MattAlbertPRB2024}. To model confinement, we introduce an external potential $V(\mathbf{r})$  and define the total Hamiltonian as
\begin{equation}
H_{\mathrm{conf}} = H_{\uparrow} + \textbf{I}_{2x2}V(\mathbf{r}),
\label{eq:confH}
\end{equation}
where $\textbf{I}_{2x2}$ is the 2x2 identity matrix. To obtain the single-particle confined states and energies, we expand the wavefunction in the basis of bulk eigenstates given in Eq.~(\ref{eq: BulkWF}). In practice, we truncate the basis to $N_B$ bulk bands, which serves as a convergence parameter for the description of states near the top of the valence band. We note that the continuum model parameters are chosen to accurately capture the highest valence bands, while the description of more remote bands is less quantitative. Since our focus is on topological edge states governed by the Chern numbers of the top most valence band, the main results are insensitive to the detailed description of deeper valence bands.

The  wavefunction for the confined system is written explicitly as
\begin{equation}
    \psi^s(\mathbf{r})
    =
    \sum_{\mathbf{k}}   \sum^{N_B}_{p=1}
    C^s_{p,\mathbf{k}}\,
    \Phi_{p,\mathbf{k}}(\mathbf{r}),
    \label{eq:ConfWF}
\end{equation}
where $s$ labels the eigenstate, and here $p$ indexes bands starting from the top most band ($p=1$). Solving the Schrodinger equation with Eq.~(\ref{eq:confH}) and Eq.~(\ref{eq:ConfWF}), we obtain the following matrix eigenvalue problem:
\begin{equation}
\epsilon_{p,\mathbf{k}}\, C^s_{p,\mathbf{k}}
+ \sum_{\mathbf{k}',p'}
C^s_{p',\mathbf{k}'}
\left\langle \Phi^{p}_{\mathbf{k}} \middle| \textbf{I}_{2x2}V(\mathbf{r}) \middle| \Phi^{p'}_{\mathbf{k}'} \right\rangle
=
E_s \, C^s_{p,\mathbf{k}} \, .
\end{equation}
At this stage, we specify the confinement potential $V(\mathbf{r})$ to model a nanoribbon geometry. We impose hard-wall confinement along the direction perpendicular to the ribbon axis, which we take to be along the moiré lattice vector $\mathbf{a}_1$. Let  $\hat{\mathbf{u}}$  denote the vector \textit{parallel} to $\mathbf{a}_1$ and let $\hat{\mathbf{v}}$ denote the unit vector \textit{perpendicular} to $\mathbf{a}_1$, and define the transverse coordinates $ u = \mathbf{r} \cdot \hat{\mathbf{u}}$,  $v = \mathbf{r} \cdot \hat{\mathbf{v}}$

The hard-wall potential of width $W$ is then
\begin{equation}
V(\mathbf{r}) \equiv V(v) =
\begin{cases}
0, & |v| \leq \dfrac{W}{2}, \\
V_0, & |v| > \dfrac{W}{2},
\end{cases}
\label{eq:HardWall}
\end{equation}
and take the limit $V_0 \to -\infty$, such that the wavefunction vanishes at $v = \pm W/2$. The hard-wall limit is implemented numerically by taking $V_0$ to be a large negative energy, appropriate for hole confinement; in the calculations below we use $V_0=-10^5$ meV and verify convergence with respect to increasing $|V_0|$. This choice confines carriers to a ribbon of width $W$ oriented along $\mathbf{a}_1$, while preserving translational invariance along the longitudinal direction. As a consequence, the crystal momentum along $\mathbf{a}_1$ ($k_u$) remains a good quantum number. The ribbon width $W$ determines how the hard-wall boundary intersects the moiré lattice and therefore fixes the effective edge termination. Due to the choice of $\mathbf{a}_1$ as the periodic direction, straight hard-wall cuts of the moiré lattice naturally realize zigzag-type terminations, whereas armchair-type terminations would require boundaries not aligned with the lattice vectors. The specific choice we take is $W = 12\,\mathbf{a}_2 \cdot \hat{\mathbf{v}}$ places the boundary  between two neighboring moiré sites, this is seen in Fig.~\ref{fig:fig1}(b) (not to scale) as the hardwall boundary cuts in the middle of the blue sites shown, while the holes localize on the bright yellow regions. 

We now compute the matrix elements with respect to our hard wall confinement. They are given as 
\begin{equation}
\begin{aligned}
\left\langle \Phi^{p}_{\mathbf{k}} \middle|\textbf{I}_{2x2} V(v) \middle| \Phi^{p'}_{\mathbf{k}'} \right\rangle
&= \frac{1}{N_2 a_M}
\sum_{\mathbf{G},\mathbf{G}'} \sum_{\alpha=1,2}
\delta_{G_u,G'_u}\, \delta_{k_u,k'_u} \\
&\quad \times
A^{\alpha *}_{p,\mathbf{k}+\mathbf{G}}\,
A^{\alpha}_{p',\mathbf{k}'+\mathbf{G}'}\,
\mathcal{F}\!\left(k_v,G_v,k'_v,G'_v\right),
\end{aligned}
\label{eq:CouplingMatrixElements}
\end{equation}

where 
\begin{equation}
\mathcal{F}\left(k_v,G_v,k'_v,G'_v\right)=\frac{2V_0}{\Delta Q}\left[\sin{\frac{\Delta QL_v}{2}}-\sin{\frac{\Delta QW}{2}}\right]
\end{equation}
 is the Fourier transform  of the hard wall potential. We have defined $\Delta Q=k_v+G_v-k'_v-G'_v$ for compactness, where $k_v$ and $G_v$ denote the components of $\mathbf{k}=(k_u,k_v)$ and $\mathbf{G}=(G_u,G_v)$ along the $\hat{\textbf{v}}$ direction. The length $L_v=N_2\textbf{a}_2\cdot\hat{\textbf{v}}$ corresponds to the system size along the longitudinal direction. It is apparent from Eq.~(\ref{eq:CouplingMatrixElements}) that we couple bulk states whose vectors along the $v$ direction differ, retaining $k_u$ as a good quantum number. It is convenient to rewrite the wavefunction in Eq.~(\ref{eq:ConfWF}) as
 \begin{equation}
       \psi^s_{k_u}(\mathbf{r})
    =
    \sum_{k_v}   \sum^{N_B}_{p=1}
    C^s_{p,k_u,k_v}\,
    \Phi_{p,k_u,k_v}(\mathbf{r}).
    \label{eq:confWFku}
 \end{equation}
We present several numerical results; convergence and implementation details are given in App.~\ref{App:NumericalDetails}.
Fig.~\ref{fig:fig2} shows the band structure of the moiré nanoribbon for four representative twist angles. To clarify the associated bulk gap closings, Fig.~\ref{fig:fig3} shows the energy gaps between the first and second bands, $\epsilon_{12}=\min(E_1)-\max(E_2)$, and between the second and third bands, $\epsilon_{23}=\min(E_2)-\max(E_3)$, for the corresponding single-valley bulk continuum model as a function of twist angle. The angle $\theta = 1.43^\circ$ corresponds to the magic angle of the continuum model, where the band flattening is most pronounced. Near $\theta \sim 1.5^\circ$ a gap closure occurs between the second and third bulk bands\cite{Devakul2021Twisted}, thus we take $\theta = 1.67^\circ$, shortly after this gap closure. To illustrate the evolution of the spectrum away from the magic-angle regime, we further include $\theta = 2.5^\circ$ as an intermediate angle. Finally, $\theta=3.5^\circ$ is chosen as a commensurate angle above the gap-closing point between the first and second bands near $\theta\approx3.3^\circ$, marking a qualitative change in the low-energy band structure. In the bulk continuum model at $\theta = 1.43^\circ$, the topmost band carries a Chern number $+1$, while the second band has Chern number $-1$. As the twist angle increases, a gap closing occurs between the second and third bands leading to a change of the Chern number of the second band from $-1$ to $+1$; however, the cumulative Chern number of the bands below the first gap remains unchanged. At larger twist angles, $\theta \gtrsim 3.3^\circ$, the gap between the first and second bands closes. Beyond this point, there is no longer a well-defined gap between the first and second band, so the corresponding edge branch is no longer spectrally isolated as an in-gap edge mode.

To gain further insight into the nature of the in-gap edge states forming between the first and second bulk bands, we focus on the magic-angle at $\theta=1.43^\circ$. In this regime, the nanoribbon edge dispersion consists of two counter-propagating branches connecting the bulk bands, as dictated by  the bulk–edge correspondence. We label these branches according to the sign of their group velocity, $v_g\sim\partial E(k)/\partial k$, and refer to them as branch 1 (left-moving) and branch 2 (right-moving), as indicated in the insets of Fig.~\ref{fig:fig4}. For each branch, we select the corresponding eigenstates $\psi^{s}_{k_u}(\textbf{r})$ within the gap between the first and second bulk bands and compute their real-space charge density. We define the total charge density of branch 1 and 2 as
\begin{equation}
    \rho^{1,2}(\textbf{r})
=\frac{1}{N_1}\sum_{k_u}
\big| \psi^{1,2}_{k_u}(\mathbf{r})\big|^2.
\end{equation}
where the sum over $k_u$ is restricted according to the grayed points in the inset of Fig.~\ref{fig:fig4}.
To further visualize the spatial structure of the edge states, it is convenient to look at slices along the finite direction. Hence, we average over the periodic direction:
\begin{equation}
    \bar{\rho}^{1,2}(v)
    =
    \int du \, \rho^{1,2}(\mathbf r),
    \qquad \mathbf r = (u,v).
    \label{eq:IntegratedDensity}
\end{equation}
Fig.~\ref{fig:fig4} (a) and (b) examine the real-space structure of the in-gap edge modes. For each branch, we compute the layer-resolved charge density $\rho$, integrated across the ribbon edge, as a function of position along the width. The resulting profiles show that the edge modes are strongly localized at the boundary and exhibit pronounced layer polarization that depends on the propagation direction, with right- and left-moving branches dominated by opposite layers. This layer polarization can be interpreted as a pseudospin polarization associated with the layer degree of freedom in the continuum model, with opposite pseudospin character for counter-propagating edge modes. Also noteworthy, the localization length is extremely short: the edge states are essentially confined to a single moiré site and show negligible weight on the second site, as confirmed by the real-space charge densities in Fig.\ref{fig:fig4}. This strong confinement and pseudospin polarization implies highly robust edge modes that are insensitive to the ribbon width and exhibit minimal hybridization across opposite edges.

We note that the edge modes shown here correspond to a single spin/valley sector of the continuum model and therefore appear as chiral edge states in the nanoribbon geometry. If both spin sectors were included, time-reversal symmetry would be restored at the single-particle level, and the edge spectrum would consist of helical pairs of counter-propagating modes related by time reversal.  In the present work, however, we focus on a single spin sector motivated by the strong interaction effects reported in moiré TMD homobilayers, such as $\mathrm{WSe_2}$ and $\mathrm{MoTe_2}$ at relatively small twist angles, where, at half-filling of the top valence band, electron–electron interactions can drive spontaneous spin/valley polarization\cite{Devakul2021Twisted,abouelkomsan2024band,zhang2021electronic}. As such, one realizes a Haldane quantum anomalous Hall insulator\cite{Wu20219ContinuumModel,Devakul2021Twisted}. In this regime, the low-energy physics is effectively captured by a single spin/valley component.

\section{Displacement-field control of moir\'e edge modes at the magic angle}
\begin{figure}[t]
    \centering
    \includegraphics[width=\linewidth]{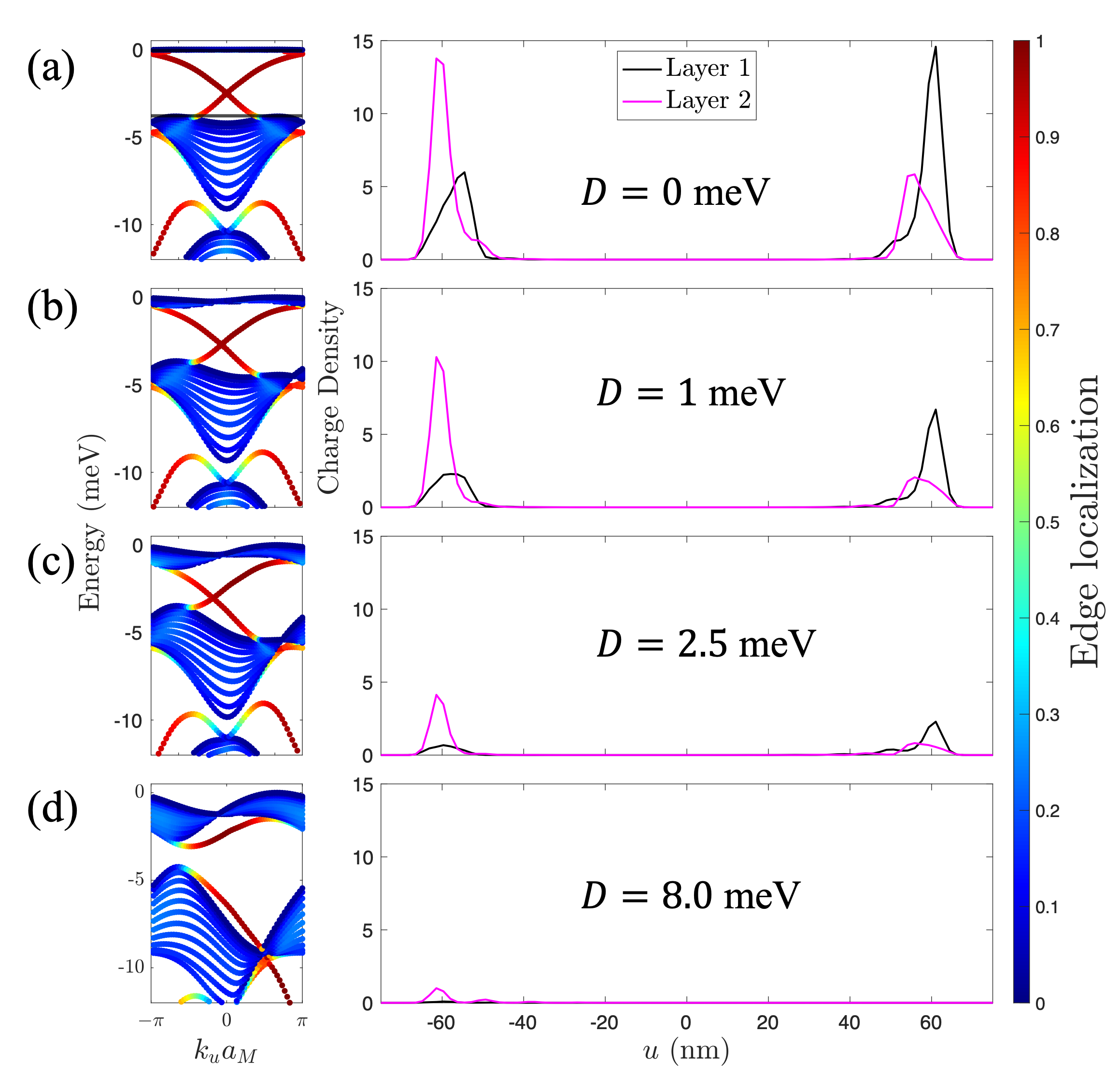}
    \caption{
    Edge-state spectra of a moiré nanoribbon at the magic angle $\theta = 1.43^\circ$ for perpendicular displacement fields with $D = 0.0, 1.0$,$ 2.5$ , and $8.0$ meV, shown from (a-d) respectively. The color scale on the far right encodes the degree of edge localization of each eigenstate. Layer-resolved edge charge density integrated across the ribbon edge for each field strength is shown in the right panel with black (magenta) curves denoting contributions from layer 1 (layer 2).
  }
    \label{fig:fig5}
\end{figure}
\label{sec:efield}

An external displacement field provides a direct and experimentally accessible knob to manipulate moir\'e edge modes by modifying the layer-resolved onsite energies and, consequently, the layer polarization and spatial confinement of the in-gap states. In twisted transition-metal dichalcogenide homobilayers, perpendicular displacement fields are known to strongly affect the topological properties of the bulk moir\'e bands by tuning interlayer hybridization and breaking inversion symmetry. In particular, sufficiently strong displacement fields can drive topological phase transitions characterized by changes in the Chern numbers of the isolated moir\'e bands, thereby controlling the existence and chirality of the associated edge states~\cite{Devakul2021Twisted, Wu20219ContinuumModel}.  

While these bulk transitions have been extensively studied, much less is known about how moderate displacement fields affect the real-space structure of topological edge modes in finite moir\'e systems. In a nanoribbon geometry, a displacement field can continuously reshape the edge-state wavefunctions, altering their layer polarization (pseudospin), spatial width, and degree of hybridization with bulk states without necessarily closing the bulk gap. This provides a powerful route for manipulating topological edge modes beyond simple on--off control via bulk gap closings. In the following, we investigate how an external perpendicular displacement field modifies the in-gap edge states in twisted $\mathrm{WSe_2}$ nanoribbons at the magic angle.

We incorporate the field at the single-particle level by adding a layer-dependent electrostatic potential to the Hamiltonian  in Eq.~(\ref{eq:confH}),
\begin{equation}
H_D=H_{\rm conf}+V_D,
\end{equation}
with
\begin{equation}
V_D=\frac{1}{2}D\sigma_z,
\label{eq:HE}
\end{equation}

where $\sigma_z$ is the Pauli matrix acting on the layer (pseudospin) degree of freedom and $D$ denotes the applied perpendicular displacement field. In analyzing the resulting edge states, two competing effects must be considered. First, increasing $D$ can drive the system toward a closing of the bulk gap between the first (top of the VB) and second band, inducing a topological phase transition. Second, the displacement field introduces a layer-dependent potential that progressively polarizes the electronic states toward one layer, thereby modifying the layer composition and spatial confinement of the edge modes.

Figure~\ref{fig:fig5} shows the edge-state spectra of the moir\'e nanoribbon for different displacement-field strengths, together with the corresponding layer-resolved edge charge densities. The in-gap edge states are identified from the ribbon spectrum and their real-space density is computed as
\begin{equation}
    \rho^{\rm edge}(\mathbf{r})
    = \frac{1}{N_1}
      \sum_{s,\,k_u \in E_{\rm gap}}
      \big| \psi^{s}_{k_u}(\mathbf{r}) \big|^2 ,
\end{equation}
where the sum runs over all eigenstates whose energies lie within the bulk gap.

At zero displacement field, the edge-state charge distribution reflects the chiral nature of the topological modes discussed above and shown in Fig.~\ref{fig:fig4}, with opposite edges predominantly associated with different layers. Upon applying a weak displacement field, charge is progressively transferred from layer 1 to layer 2, leading to enhanced accumulation near one edge of the ribbon, as seen in the right panels of Fig.~\ref{fig:fig5}(b) and (c). 

Simultaneously, the displacement field reduces the bulk gap, thereby promoting hybridization between edge and bulk-like states. This results in a gradual depletion of charge from the boundary and an increased redistribution into the bulk. 

For sufficiently large electric fields, the bulk bands cross and the system undergoes a transition into a topologically trivial regime (Fig.~\ref{fig:fig5}(d)). In this regime, the edge charge is strongly suppressed and becomes purely layer-polarized. The remaining in-gap states originate from the zigzag edge termination rather than from a difference in Chern numbers between the top two bands, analogous to the well-known edge states in graphene nanoribbons~\cite{nakada1996edge}.

\section{The Haldane Model} 
\label{sec:Haldane}
\begin{figure}[t]
    \centering
    \includegraphics[width=\linewidth]{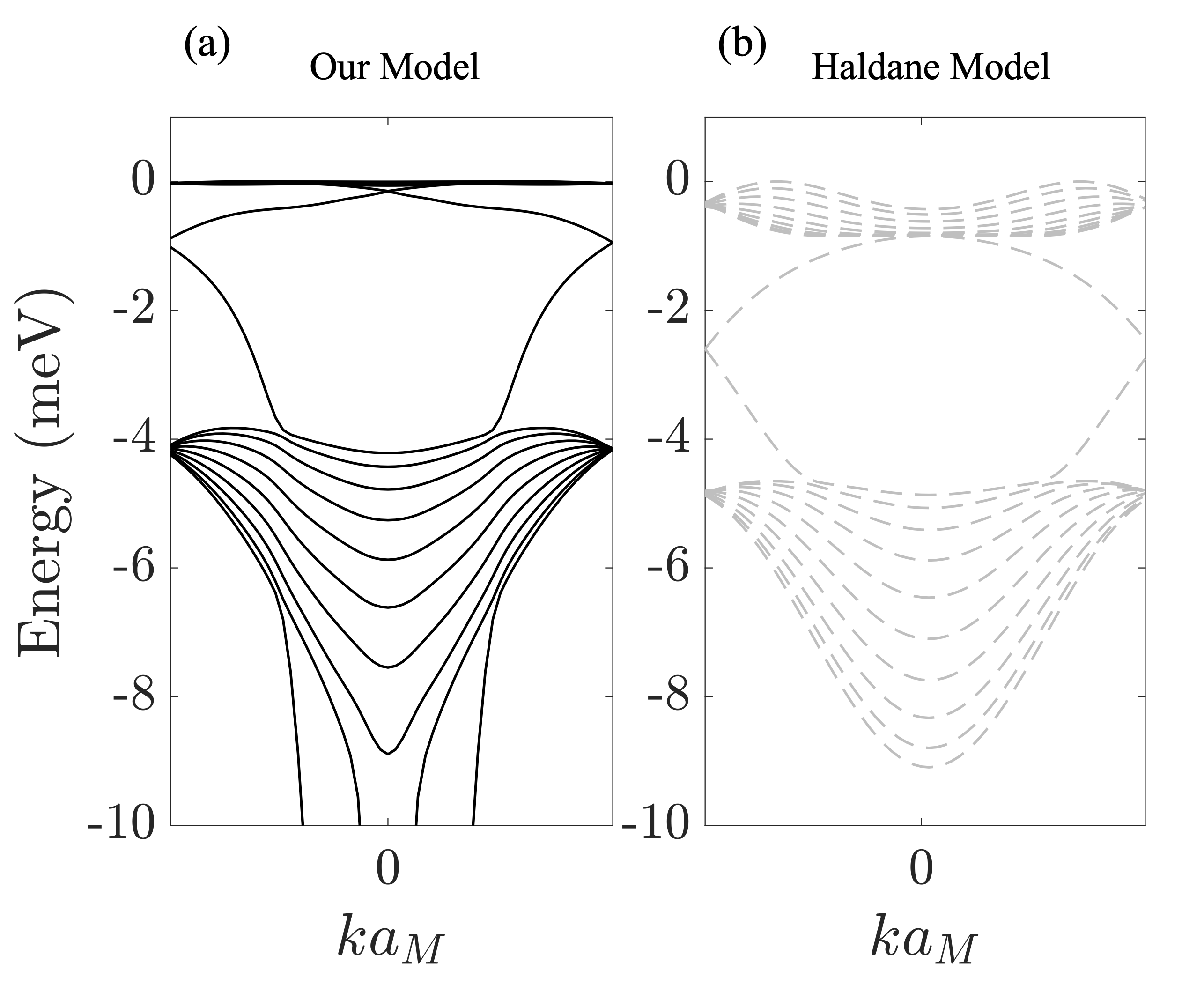}
    \caption{
     Band structure of a moiré nanoribbon at the magic angle $\theta = 1.43^\circ$ for (a) our model with $N_B=2$ and (b) the Haldane model. The parameters used for the Haldane model are taken from Fig.4(b) of Ref.~[\onlinecite{Devakul2021Twisted}] up to fourth neighbors.
  }
    \label{fig:fig6}
\end{figure}

The topological properties of the moiré bands in twisted $\mathrm{WSe_2}$ admit a simple and physically transparent interpretation in terms of an effective tight-binding model. This connection was first described by Wu \textit{et al.}~[\onlinecite{Wu20219ContinuumModel}], who showed at small-twist angles that the top valence moiré bands for $\mathrm{MoTe_2}$ homobilayers for one spin type can be mapped onto a Haldane model (or in the case of both spins a Kane-Mele model\cite{kane2005quantum,kane2005z}) with complex next-nearest neighbor hopping amplitudes. This was later done by Devakul \textit{et al.}~[\onlinecite{Devakul2021Twisted}] for t$\mathrm{WSe_2}$. In particular, the two highest valence moiré bands at the magic angle t$\mathrm{WSe_2}$ can be mapped onto a Haldane model defined on an emergent hexagonal lattice formed by the moiré potential minima\cite{Devakul2021Twisted}:
\begin{equation}
    H_h= t\sum_{\langle i,j\rangle}c^\dagger_{i}c_j+|t_2|\sum_{\langle\langle i,j\rangle\rangle}e^{i\phi\nu_{ij}}c^\dagger_{i}c_j+\cdots,
\end{equation}
where the hopping parameters we take are extracted from Fig.4(b) of Ref.~[\onlinecite{Devakul2021Twisted}]. This finite-range Haldane model should be viewed as an approximate representation of the continuum bands: it captures the topology and qualitative edge dispersion, but not the precise bandwidth of the continuum ribbon spectrum near the magic angle, where longer-range hoppings become increasingly important.

To facilitate a direct comparison with the effective Haldane description, in Fig.~\ref{fig:fig6} we restrict our continuum ribbon construction to $N_B=2$ bulk bands. In the full calculations presented above we use $N_B=9$ bands, but retaining only two bands allows for a more transparent comparison with the two-band Wannierized Haldane model.

We analyze the fully converged ribbon spectrum at the magic angle, shown in Fig.~\ref{fig:fig2}(a). To interpret these results, it is instructive to compare our continuum ribbon construction with the effective Haldane model, where edge states are typically defined for ribbons with well-specified lattice terminations, such as zigzag edges. In contrast, in our model the ribbon is generated by a confining potential in continuous space, rather than by terminating a discrete lattice, leading to a different notion of edge termination. We find that the existence and chirality of the edge modes are already captured at the two-band level. However, their detailed dispersion exhibits quantitative differences, reflecting the influence of remote bands, which are naturally included in the continuum description. We further note that the Haldane model is inherently limited by its two-band structure, for which the sum of the Chern numbers must vanish, implying that the two bands necessarily carry opposite Chern numbers. As a result, it cannot capture situations in which multiple bands contribute nontrivially to the topology in a more general multiband setting. In contrast, the continuum approach employed here naturally incorporates additional bands and is not restricted to a two-band description, allowing for a more complete characterization of the band topology.

\section{Discussion}
\label{sec:Disc}
 The results presented here establish a direct connection between continuum descriptions of moiré systems and the physics of edge states in finite geometries. By working entirely within the continuum framework, our approach circumvents the need for Wannierization and avoids the constraints associated with constructing localized lattice models for topological bands. This enables a direct and systematic treatment of boundary effects while retaining the full structure of the underlying continuum Hamiltonian. The accuracy of the model depends strictly on the limited accuracy of the bulk model used, and improvements to the continuum model would improve results pertained to the method outlined here.  Improvements have already been established in $\mathrm{MoTe_2}$, for example,  where including second harmonic terms improved the accuracy of remote bands\cite{jia2024moire}. 
 
A key feature of the edge states in the ribbon geometry is their pronounced layer (pseudospin) polarization, with counter-propagating modes associated with opposite layers. This provides a natural mechanism for controlling edge-state properties via a perpendicular displacement field, which couples directly to the pseudospin degree of freedom. As a result, the electric field enables continuous tuning of the edge-state structure, including their localization, layer composition, and hybridization with bulk states, as well as field-induced band crossings and topological transitions.

A comparison with an effective Haldane model provides a useful low-energy interpretation of the continuum results, capturing the existence and chirality of the edge modes. However, it is important to emphasize that such a mapping is valid only within a restricted parameter regime and relies on an approximate reduction to a two-band description. In realistic continuum models for twisted WSe$_2$ and MoTe$_2$, different parameter sets have been proposed in the literature, and multi-band effects can play a significant role. This further highlights the importance of the full continuum approach employed here, which naturally incorporates multi-band physics and avoids reliance on Wannierization.

Finally, we emphasize that the present continuum calculation resolves the termination of the moir\'e pattern, but not the microscopic atomic termination. It therefore describes moir\'e-defined edge states rather than additional dangling-bond or passivation-dependent atomic edge states. This separation of scales is supported by previous work on twisted graphene nanoribbons, where moir\'e edge states were found to be distinct from ordinary zigzag edge states and robust against changes in atomic edge geometry and roughness~\cite{fleischmann2018moire}. Comparisons with DFT-parametrized tight-binding models~\cite{vitale2021flat} would be valuable for capturing atomistic aspects not included here and benchmarking continuum model results.

An important direction for future work is the extension of the present framework to more experimentally realistic confinement profiles, such as smooth electrostatic potentials defining lateral quantum dots. In contrast to the hard-wall geometries considered here, soft confinement is expected to predominantly involve carriers from the topmost moir\'e band, where the continuum description is most reliable. In this regime, the low-energy physics may be effectively captured by a single isolated band, providing a controlled setting for studying confined states. The ability to treat such geometries within a continuum framework opens the door to modeling laterally gated moir\'e devices, where confinement, topology, and layer pseudospin can be tuned. This suggests a direct connection to ongoing experimental efforts in bilayer graphene, TMDs, and semiconducting heterostructures~\cite{kurzmann2019excited,eich2018spin,boddison2025valley,thureja2022electrically,lee2026revealing,inbar2023quantum}, and provides a pathway toward realizing and controlling topological quantum dots in twisted TMD systems. For more general device geometries and transport calculations, an additional practical limitation is that the projected bulk-state basis is not local in real space, therefore making direct NEGF implementations less immediate than in localized tight-binding or Wannier approaches. Extending the present framework toward transport calculations within a continuum model~\cite{tworzydlo2006sub,novik2010signatures} is therefore an interesting direction for future work.

\vspace{0.8em}\noindent  \textbf{Data availability}

\noindent{The data that support the findings of this study are available within the paper and the Supplementary Information. Other relevant data are available from the corresponding authors upon request.} 

\vspace{0.8em}\noindent \textbf{Code availability}

\vspace{0.8em}\noindent  \textbf{Acknowledgements}

\noindent{We acknowledge financial support by the Deutsche Forschungsgemeinschaft (DFG, German Research Foundation) through the Würzburg-Dresden Cluster of Excellence ctd.qmat – Complexity, Topology and Dynamics in Quantum Matter (EXC 2147, project-id 390858490).}

\vspace{0.8em}\noindent  \textbf{Author Contributions}

\vspace{0.8em}\noindent  \textbf{Conflict of Interest}
\noindent The authors declare no conflict of interest.

\vspace{0.8em}\noindent  \textbf{Corresponding Author}

\noindent Yasser Saleem: yassersaleem461@gmail.com


\appendix
\section{Numerical Details and Convergence}
\label{App:NumericalDetails}
\begin{figure*}[t]
    \centering
    \includegraphics[width=\linewidth]{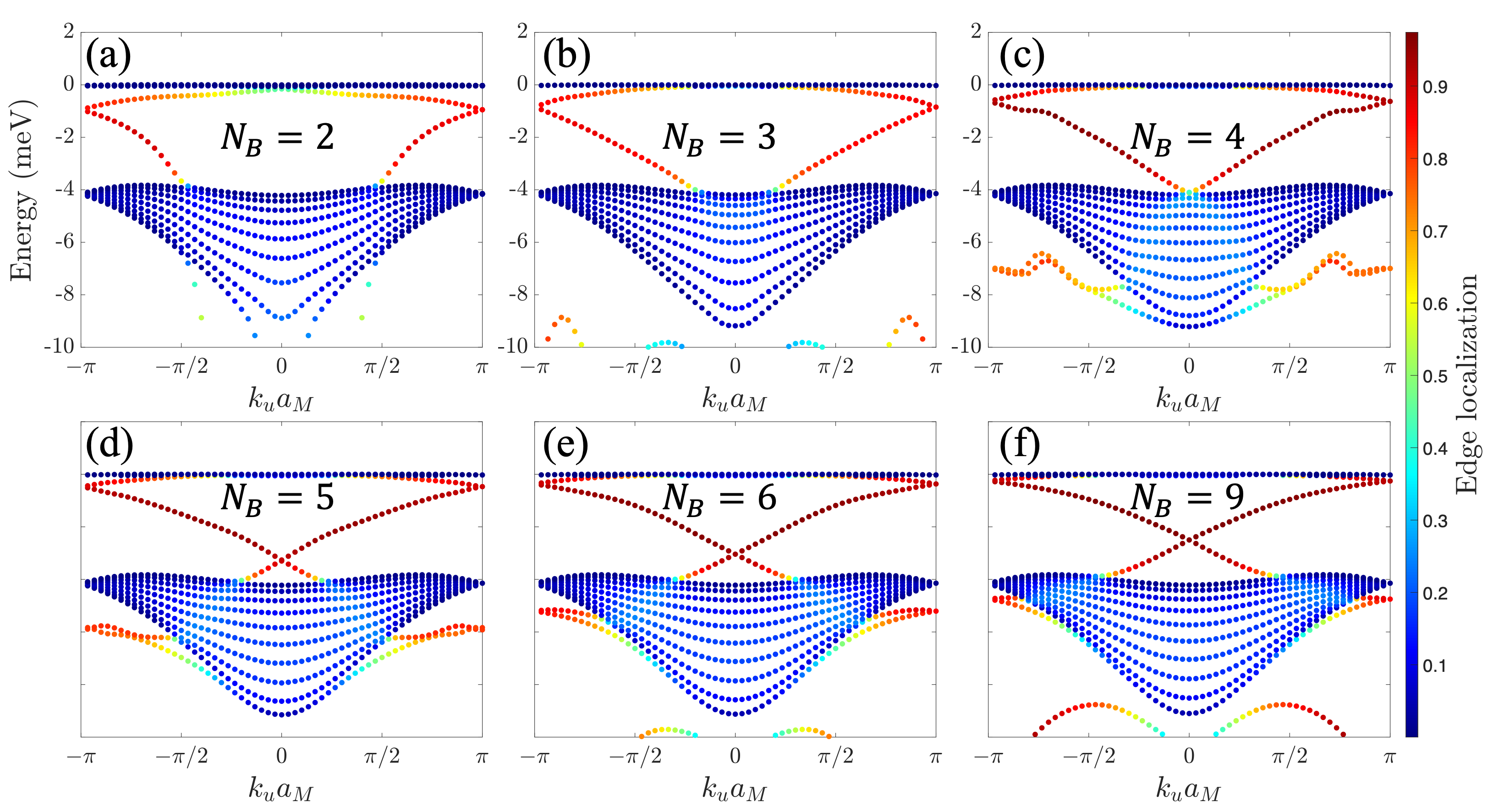}
\caption{Convergence of the moiré nanoribbon spectrum at the magic angle $\theta = 1.43^\circ$ with respect to the number of bulk bands $N_B$. Panels (a)–(f) show the edge-resolved band structure for $N_B = 2, 3, 4, 5, 6,$ and $9$, respectively.  The color scale encodes the degree of edge localization of each eigenstate. 
    }\label{fig:Appfig1}
\end{figure*}

\begin{figure*}[t]
    \centering
    \includegraphics[width=\linewidth]{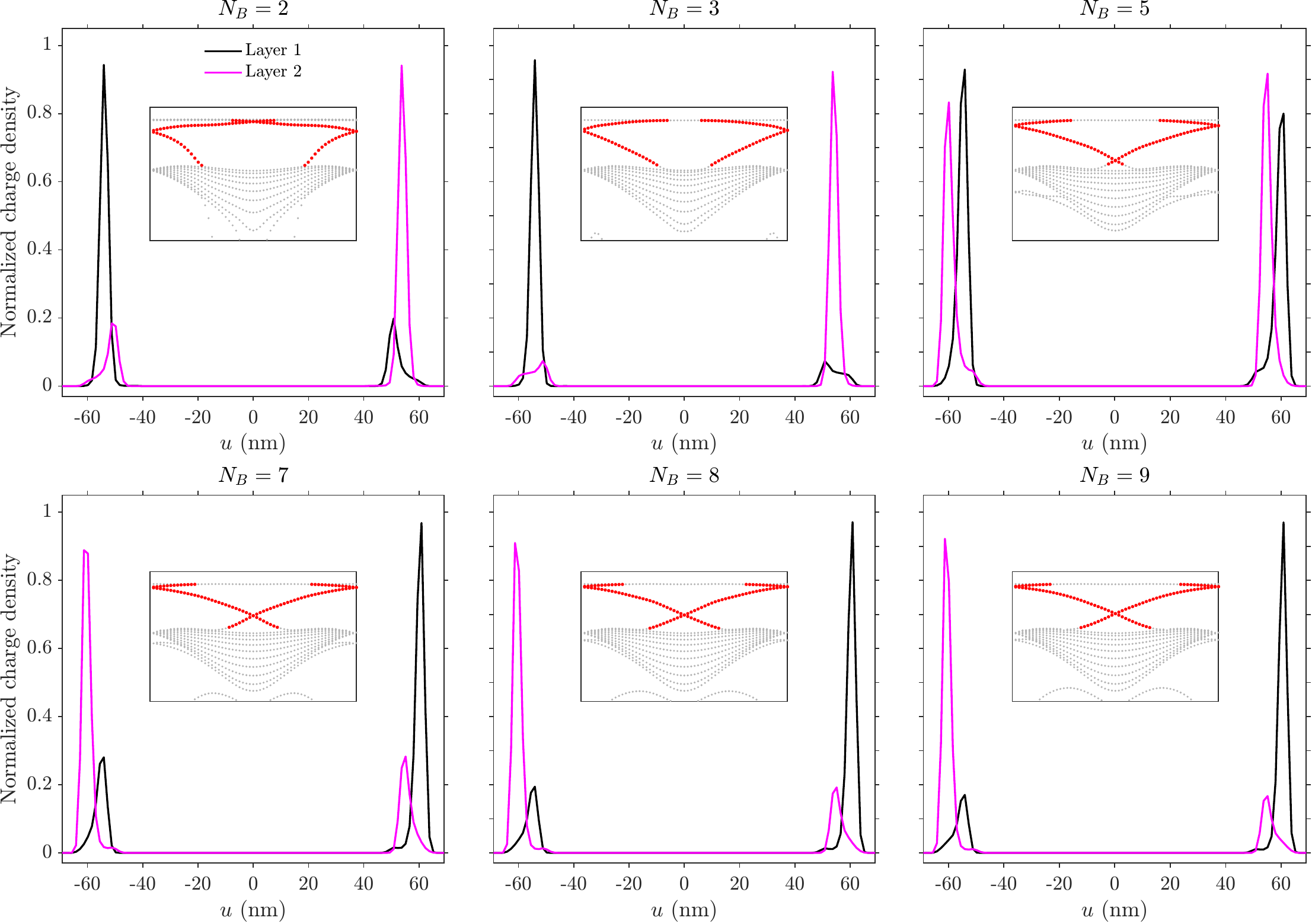}
\caption{Edge-state charge densities at the magic angle $\theta = 1.43^\circ$ for different numbers of bulk bands $N_B$. The insets show the nanoribbon spectrum for each case, with the selected in-gap edge states highlighted in red.}\label{fig:Appfig2}
\end{figure*}

Here we provide additional details regarding the numerical implementation and convergence of our results with respect to the truncation parameters used in the ribbon construction. Throughout the text we take $N_1=60$ which defines the number of $k_u$ points taken. We take $N_2=30$ which defines the number of wavevectors $k_v$. This also corresponds to a computational box substantially larger than  ribbon width. We take $N_G=61$ $\textbf{G}$ vectors which easily converges both the confined spectra and the bulk spectra, and in the main text we take $N_B=9$. See Fig.~\ref{fig:Appfig1} for details of the quantitative effects of including more bands. Figure~\ref{fig:Appfig2} shows the corresponding layer-resolved charge density of the in-gap edge states. The states remain localized near the ribbon boundaries for all values of $N_B$ considered. However, the layer-resolved structure is more sensitive to the basis size, showing that higher bands can affect the internal layer texture of the edge modes even when their spatial localization is already captured.

\section{Dependence on Moir\'e Boundary Position}
\label{App:DifferentCuts}
\begin{figure*}[t]
    \centering
    \includegraphics[width=\linewidth]{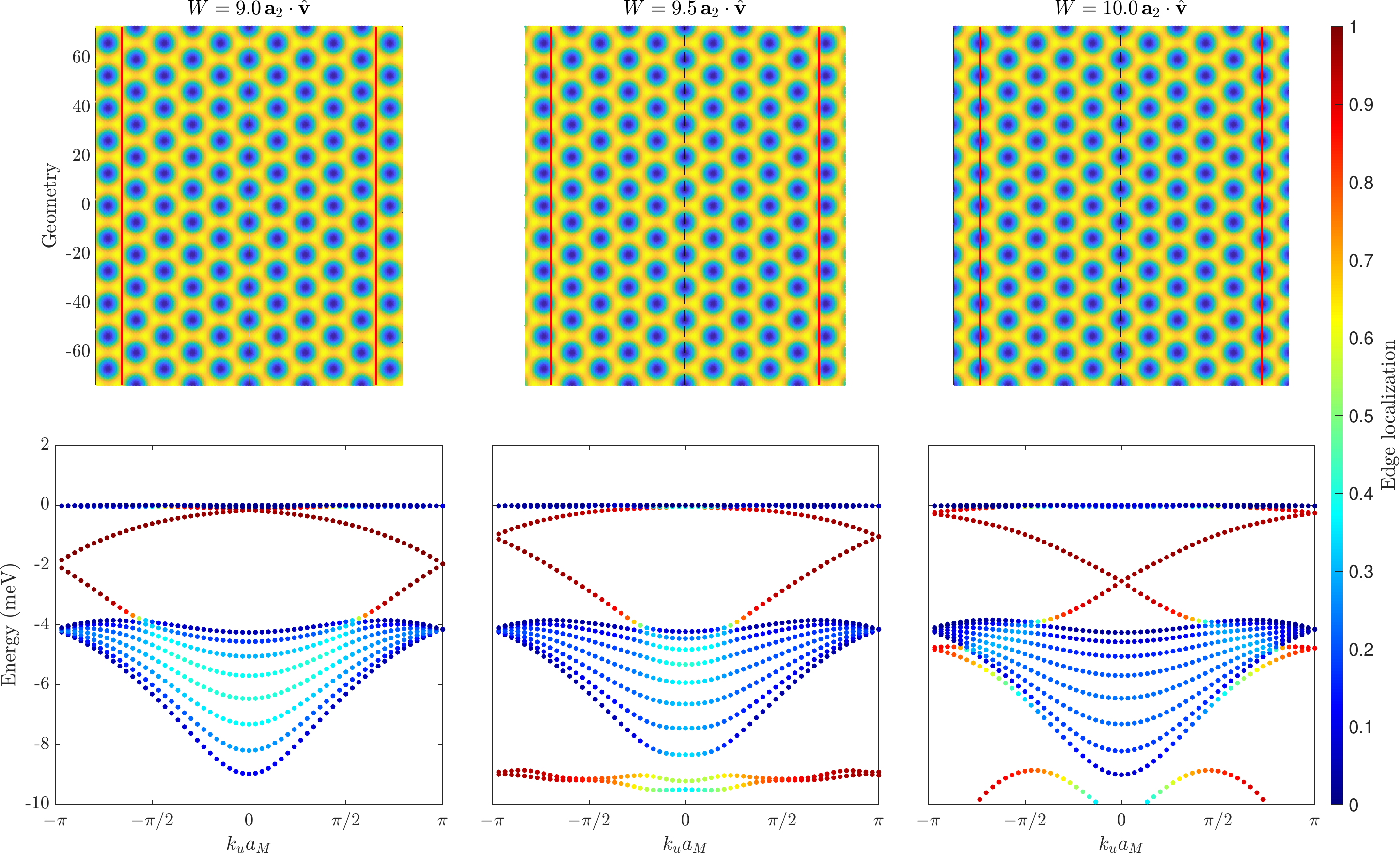}
\caption{Dependence of the moir\'e edge spectrum on the position of the hard-wall boundary. 
The top panels show the moir\'e potential together with representative ribbon terminations for different widths/cut positions, while the bottom panels show the corresponding ribbon spectra.
    }\label{fig:Appfig3}
\end{figure*}

To test the sensitivity of the edge spectrum to the position of the hard-wall boundary, we repeat the ribbon calculation for several representative cuts through the moir\'e unit cell. As shown in Fig.~\ref{fig:Appfig3}, changing the cut position modifies quantitative features of the edge dispersion, such as the detailed bandwidth and crossing momentum. However, the gap-traversing edge modes remain present for the representative cuts considered here, consistent with their moir\'e-scale topological origin.

\clearpage
\newpage
\bibliography{Main}

\clearpage
\newpage

\end{document}